\documentclass[intlimits,twoside,a4paper]{article}
\usepackage{amsmath,amssymb}
\usepackage{graphicx}
\usepackage[T2A]{fontenc}
\usepackage[cp1251]{inputenc}
\usepackage[percent]{overpic}

\usepackage[eqsecnum]{cmpj2}

\issue{2017}{20}{1}{13602}
\doinumber{10.5488/CMP.20.13602}

\title[Two-step percolation in aggregating systems]
{Two-step percolation in aggregating systems\thanks{This work is dedicated to the 60th birthday of Professor Yurij Holovatch.
}}

\author[N.~Lebovka  \textsl{et al.}]
{
N.~Lebovka\refaddr{label1}\footnote{Corresponding author, E-mail:~lebovka@gmail.com.}\,, 
L.~Bulavin\refaddr{label2}, 
V.~Kovalchuk\refaddr{label2}, 
I.~Melnyk\refaddr{label2},  
K.~Repnin\refaddr{label2}

\addresses{
\addr{label1} F.D.~Ovcharenko Institute of Biocolloidal Chemistry of the National Academy of Sciences of Ukraine,\\
42~Acad. Vernadsky Blvd., 03142 Kyiv, Ukraine
\addr{label2} Department of Physics, Taras Shevchenko Kyiv National University,
2~Acad. Glushkov Ave., 03127 Kyiv, Ukraine
}}

\authorcopyright{N.~Lebovka, L.~Bulavin, V.~Kovalchuk, I.~Melnyk, K.~Repnin, 2017}
\date{Received December 3, 2016, in final form February 5, 2017}

\begin{document}
\maketitle
\begin{abstract}
The two-step percolation behavior in aggregating systems was studied both experimentally and by means of Monte Carlo (MC) simulations. In experimental studies, the electrical conductivity, $\sigma$, of colloidal suspension of multiwalled
carbon nanotubes (CNTs) in decane was measured. The suspension was submitted to mechanical de-liquoring in a planar filtration-compression conductometric cell. During de-liquoring, the distance between the measuring electrodes continuously decreased and the CNT volume fraction $\varphi$ continuously increased (from $10^{-3}$ up to $\approx 0.3$\%~v/v). The two percolation thresholds at $\varphi_{1}\lesssim 10^{-3}$ and
$\varphi_{2}\approx 10^{-2}$ can reflect the interpenetration of loose CNT aggregates and percolation
across the compact conducting aggregates, respectively. The MC computational model accounted for the core-shell structure
of conducting particles or their aggregates, the tendency of a particle for aggregation, the formation of solvation
shells, and the elongated geometry of the conductometric cell. The MC studies revealed two smoothed
percolation transitions in $\sigma(\varphi)$ dependencies that correspond to the percolation through
the shells and cores, respectively. The data demonstrated a noticeable impact
of particle aggregation on anisotropy in electrical conductivity $\sigma(\varphi)$ measured along different directions
in the conductometric cell.

\keywords{multiwalled carbon nanotubes, colloidal suspensions, anisotropy of electrical conductivity, two-step percolation}
\pacs{61.48.De, 64.60.ah, 72.80.Tm, 73.50.-h, 73.61.-r}

\end{abstract}

\section{\label{sec:introduction}Introduction}

Classical percolation with one sharp transition from a non-conducting to a conducting state is commonly expected for
composites filled with highly conducting particles. So far, a lot of different models and equations were proposed for a description of the
electrical conductivity behavior~\cite{Han1998,Taherian2016}.

However, in many experimental observations the percolation in composites is more complicated. The presence of two-step [double percolation (DP)],
several-step (multiple percolations) and even fuzzy (smeared) type of percolation transitions has been reported by many researchers
~\cite{Lobb1978,Sharma1980,Mikrajuddin1999,McQueen2005,McQueen2004,Sheng1982,Malarz2005,Nettelblad2003,AbbasiMoud2015,Gong2016,Gubbels1995,Sumita1992,Xiu2014,Levon1993,Zhang1998b,Lebovka2006,Zhang1998a,Zhang1998,Thongruang2002,Wen2007}.
Different mechanisms related with distribution of types of particles and types of the electrical
contacts, geometrical effects, selective distribution of a conducting particle in multi-component media (e.g.,
in polymer blends), the existence of static and kinetic network formation processes, as well as the core-shell structure of particles
may be responsible for the multiple percolation thresholds.

The superconducting DP has been experimentally observed and attributed to the
distributions of particles within the composites~\cite{Lobb1978,Sharma1980}. The DP has been described accounting for different types of electrical contacts in the material: clean contacts with fraction $\alpha$ and insulator separated contacts with fraction $1-\alpha$~\cite{Mikrajuddin1999}. This parameter $\alpha$ was a relevant function of the applied pressure and temperature, the type of a matrix, the insulating layer on the particle's surface, etc. For this model, the effective medium (EM) theory predicted the existence of the DP at $\alpha<1$. An experimental realization for polymer/filler composites with the multiple percolation thresholds has been explained accounting for the local variations in filler concentration and/or irregularities in the shape and orientation of particles~\cite{McQueen2005,McQueen2004}. The multiple percolation was attributed to the presence of various geometric shapes and correlated arrangements of particles.

For a two-dimensional (2D) square lattice, the DP has been simulated accounting for different edge-edge nearest neighbor NN contacts and next-nearest neighbor NNN (von Neumann's neighborhood) contacts~\cite{Sheng1982}. For NN (or NNN) and NN+NNN (Moore's neighborhood, composed of a central cell and eight cells that surround it) contacts, the percolation thresholds are $0.5927$ and $1-0.5927=0.4073$, respectively \cite{Malarz2005}.
The existence of the DP due to a geometrical effect has been confirmed by experiments with angular and rounded silicon
carbide (SiC) particles in a polymer rubber~\cite{Nettelblad2003}. For angular SiC grains with sharp edges and rather flat surfaces, the DP was observed. However, for the rounded SiC grains, only one threshold was observed. The value of conductivity at a plateau
between two percolations was found to be close to the geometric mean of the limiting conductivities at low ($\sigma_{\text i}$) and at high ($\sigma_{\text c}$) concentrations of particles. The concentration dependence of electrical conductivity was simulated using the three-dimensional (3D) impedance network model accounting for the presence of both edge and face contacts, where the model also revealed the presence of the DP.

A particular type of \textit{blend} double percolation (BDP) has been observed in immiscible polymer blends filled with conducting
particles~\cite{AbbasiMoud2015}. For these systems, the conducting particles have different affinity to polymer components
A and B and are capable of finely dispersing only in one of them, e.g., in A. This caused a  selective spatial localization
of conducting particles in A component. In such blends, the percolation in electrical conductivity requires both percolation of conducting particles within the component A and connectivity of component A within component B.
The BDP has been frequently observed using a carbon black (CB) conducting filler~\cite{Gong2016,
Gubbels1995,Sumita1992,Xiu2014}. A fine regulation of electrical conductivity is possible by changing the concentration of
CB in phase A and the concentration of phase A in phase B. A general concept of multiple percolation
hierarchy for CB filled polymer blends was developed~\cite{Levon1993}. The percolation concentration of CB in the
blend may be rather low~\cite{Zhang1998b}. This conclusion was supported by the data of computer simulation of the
percolation behavior of composites containing small conducting particles between large isolating
particles~\cite{Lebovka2006}. The BDP has been also observed in the polymers filled with a short carbon fiber \cite{Zhang1998a} and with Ketjenblack \cite{Zhang1998}. It was noted that carbon nanoparticles can affect the morphology of the blends. The combination of fillers (graphite
and carbon fiber) has been used to enhance the interparticle connectivity and increase the electrical conductivity~\cite{Thongruang2002}. The effects were explained in terms of \textit{bridged} DP mechanism.

An interesting example of DP in carbon fiber reinforced cement-based materials has been experimentally discovered~\cite{Wen2007}. The percolation was observed at $\approx 0.30{-}0.80$~vol.\% fibers in the paste portion and at $70{-}76$~vol.\% carbon fiber cement paste in mortar.

In recent years, the composites filled with carbon nanotubes (CNTs) have attracted great attention. Such composites have many
fascinating properties due to their versatility of applications in antistatic devices, electromagnetic interference
shielding materials, capacitors, and sensors~\cite{Soares2016}. A significant reduction in percolation concentration of CNTs due to the BDP has been reported
\cite{Al-Saleh2016,Baudouin2010,Chen2015,Gao2015,Li2015,Maiti2013,Mamunya2016,Nair2015,Nasti2016,Pang2012,Poetschke2003,Xu2015}.
The DP in CNT epoxy composites has been attributed to a static (at higher concentration) and to a kinetic network formation (at lower
concentration) processes~\cite{Kovacs2007}. The electrical conductivity plateau was attributed to the presence of a superstructure
of flocculated particles. Flocculation can noticeably affect the value of the percolation threshold. The proposed model also
accounted for the magnitude of individual interparticle contact resistance. The DP in CNT filled liquid
crystals medium has been attributed to a core-shell structure of conducting particles, where the EM theory
has been used to explain the experimental data~\cite{Tomylko2015}.

Percolation can be greatly affected by interaction between conducting particles and their aggregation. The theory of
correlated percolation predicts a strong dependence of the percolation threshold upon the details of interaction~\cite{Safran1985,
Weinrib1984}. The effects of aggregation in multiple percolation are still far from being completely understood. For well-dispersed CNTs (with small size of agglomerate) in polymeric composites, the increase in the electrical conductivity~\cite{Jamali2013} and very small percolation thresholds~\cite{Krause2016} have been observed. However, a shear-induced re-aggregation in well-dispersed systems facilitated the interconnectivity of CNTs as well as could result in a decrease of the percolation threshold~\cite{Skipa2010}. Regarding this behavior two controversial effects can be important. On the one hand, a good dispersion is helpful in forming well distributed paths. On the other hand,  partial agglomerations are helpful in reducing the distance between the tubes to the tunneling range.

The present work is devoted to experimental and computational studies of percolation in an aggregating system.
In the experimental part the electrical conductivity $\sigma$ of suspensions of multiwalled CNTs in decane was measured
using a home-made filtration-compression conductometric cell. The decane was continuously expressed from the suspension by
pressure. During de-liquoring, the concentration of CNTs increased the changes in $\sigma$ encompassed by both the effects of changes
in concentration and by agglomeration of CNTs. In the computational part, the Monte Carlo (MC) model was developed accounting for the core-shell
structure of conducting particles or their aggregates, the tendency of particle aggregation, the formation of solvation shells,
and the elongated geometry of the conductometric cell. Both experiment and the MC data revealed DP. An impact of particle aggregation on anisotropy in electrical conductivity measured along different directions
in the conductometric cell is discussed.  In section~\ref{sec:methods}, the materials, experimental methods, MC model, and details
of simulation are presented. The obtained results are discussed in section~\ref{sec:results}. Conclusions and final remarks are formulated
in section~\ref{sec:conclusion}.

\section{Materials and methods \label{sec:methods}}

\subsection{Materials}
Multiwalled CNTs, obtained by chemical vapour deposition of graphite in the gas phase with a catalyst $\text{FeAlMo}_{0.07}$
(Spetsmash, Ukraine), were used as a pristine material~\cite{Melezhik2005}. To separate the CNTs from the catalyst and mineral
impurities, the material was treated by aqueous solutions of alkali (NaOH) and hydrochloric acid (HCl). The samples
were filtered to remove the excess acid and repeatedly washed with distilled water until the pH value of distilled water was
reached. The studied CNTs have the outer diameter $d\approx 20{-}40$~nm and length $l\approx 5{-}10$ microns.
The density of CNTs was assumed to be the same as the density of pure graphite, $\rho_{n}=2.1$~g/cm$^{3}$.
Decane, $\text{CH}_{3}(\text{CH}_{2})_{8}\text{CH}_{3}$ (TU 6-09-3614-74, Ltd. ``Novohim'' Kharkiv, Ukraine) was used as a
fluidic matrix. Pure decane has a fairly small electrical conductivity ($<10^{-15}$ S/cm \cite{Guo2012}),  solubility of water in decane being also very low ($7.2\times 10^{-3}\%$ \cite{Smallwood1996}).

The CNT-decane colloidal suspensions
were obtained by adding appropriate
weights of CNTs to the decane with subsequent $20{-}30$~min sonication of the mixture using a UZD-22/44 ultrasonic disperser
(Ukrprylad, Sumy, Ukraine) at a frequency of $44$~kHz and output power of $150$~W. The measurements were started immediately after sonification.

\subsection{Filtration-compression conductometric cell}
\begin{figure}[!t] 
\centering
\includegraphics [width=0.45\linewidth] {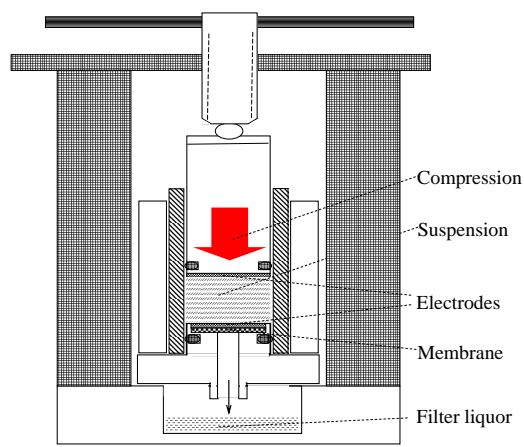}
\caption{(Color online) Sketch of a vertical planar filtration-compression conductometric cell. The meshed bottom electrode was placed onto the
membrane. The CNT-decane suspension between the upper and bottom electrodes was submitted to mechanical de-liquoring, the decane was
filtered through the membrane, and filter liquor was collected within the container.}
\label{fig1}
\end{figure}

Figure~\ref{fig1} shows the sketch of home-made vertical planar filtration-compression conductometric cell. CNT-decane suspension was
placed into calibrated cylinder between two electrodes. The electrode surfaces were covered by the plates of porous nickel.
A meshed bottom electrode permits filtration of a dispersed medium. The displacement of the upper electrodes was realized by the screw-down press and was controlled by cathetometer MK-6 (LOMO, Saint Petersburg) with a precision of $\pm0.01$~mm.
The filter liquor was collected within the bottom container. An initial volume concentration of suspension was $\varphi_{\text i}=
0.1$\%~v/v. The electrical conductivity was measured using AM~3003 conductivity meter (Data.com) at the temperature of 293~K,
frequency of 1~kHz and voltage of 0.25~V. The choice of 1~kHz made it possible to avoid the effects of near-electrode polarization
and migration of CNTs in an external electric field.
To avoid CNTs' sedimentation the conductometric cell was intensively shaked before each measurement.

\subsection{Microstructure of suspensions}
\begin{figure}[!h] 
\centering
\includegraphics [width=0.55\linewidth]{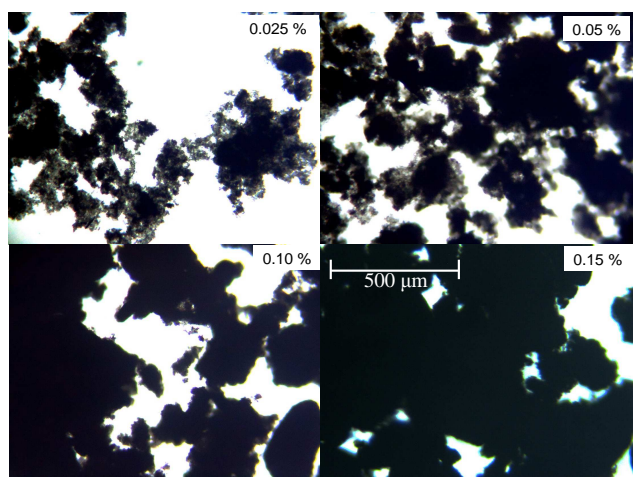}
\caption{Micro-photos of CNT-decane suspensions at different volume fractions of CNTs, $\varphi$, and temperature of 293~K.}
\label{fig2}
\end{figure}

Optical microphotographs of CNT-decane suspensions were obtained using a microscope Biolar~03-808 (Warsaw, Poland). Suspensions were placed in a flat cell with the layer thickness of 70~\textmu m. All measurements were done at 293~K. Figure~\ref{fig2} shows examples of micro-photos of CNT-decane suspensions at different CNT-decane suspensions at different volume
fractions of CNTs, $\varphi$. The CNT aggregates became visible at low concentration ($\varphi\approx 0.01\%$) and they grown in size with an  increase of $\varphi$. Finally, a large spanning aggregate with the size exceeding the microscope visual field was formed at $\varphi\approx 0.025\%$.

\subsection{Monte Carlo computational model}
The MC simulation was used to imitate the changes of electrical conductivity measured in the planar
filtration-compression cell during mechanical de-liquoring of colloidal suspension.
The tendency of particle aggregation was accounted for using the previously described model of the interactive
cluster-growth~\cite{Anderson1988,Lebovka2015}. A 2D model on a square lattice was used. All sites were initially empty and have small electrical conductivity, $\sigma_{\text i}(=1)$. The empty sites were randomly filled with
conducting particles of much larger electrical conductivity, $\sigma_{\text c}(=10^{6})$.

The core-shell structure of conducting particles was assumed. The cores of conducting particles were covered with a conducting shell of
intermediate electrical conductivity  $\sigma_{\text s}(=10^{3})$. The shells can account for the presence of solvation shells
around the conducting species of a smaller electrical conductivity.

\begin{figure}[!t] 
\centering
\includegraphics [width=0.85\linewidth] {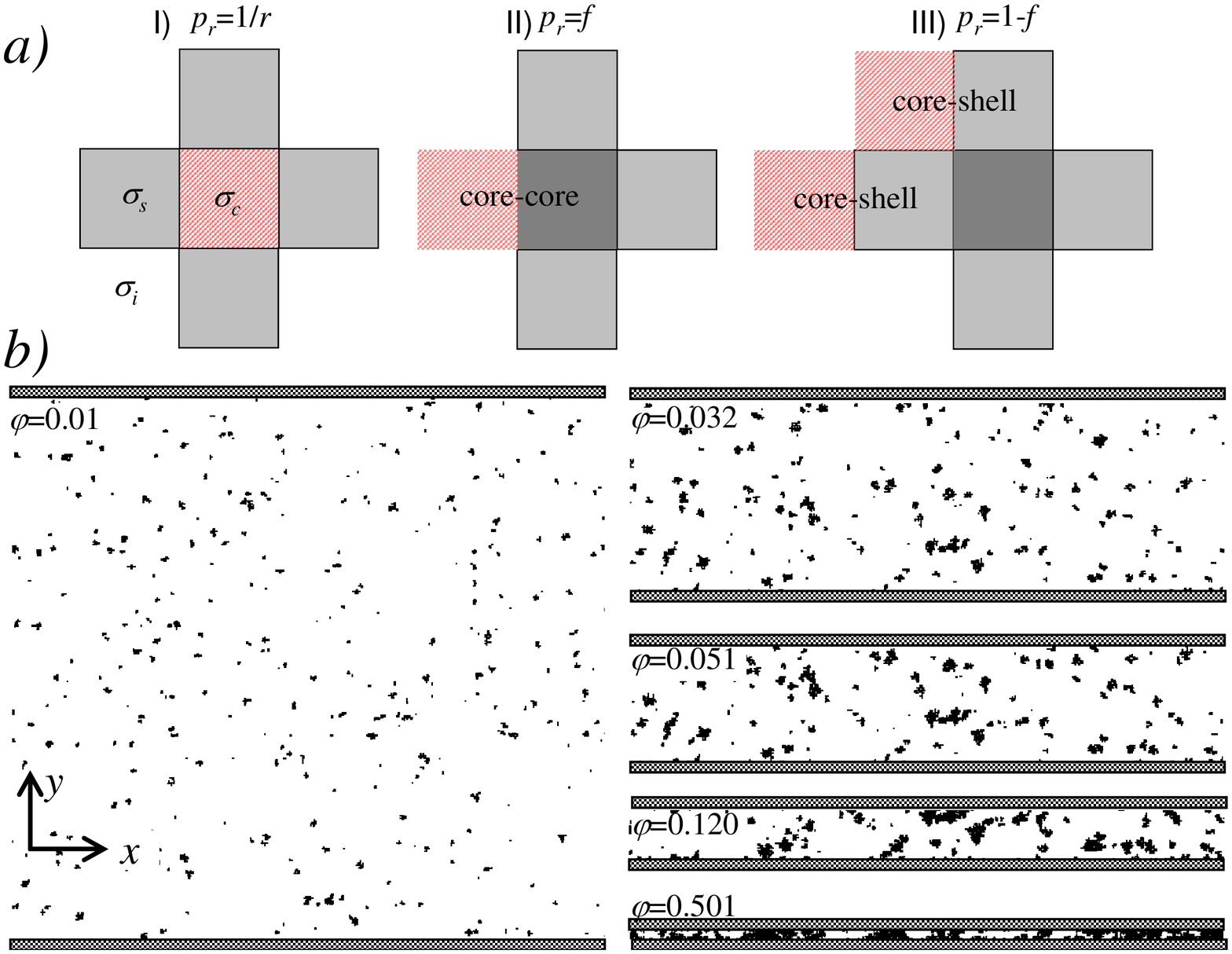}
\caption{(Color online) Description of the aggregation model of core-shell particles (a) and examples of simulated aggregation patterns during the filtration-compression process at different volume fractions of conducting particles $\varphi$ (b).
\label{fig3}}
\end{figure}

The probability of filling a new empty site $p_{r}$ was dependent on the state of the NNs in the
Moore neighborhood (it is composed of a central cell and eight cells that surround it):
\begin{itemize}
  \item In the absence of direct contacts (core-core or core-shell) the probability of site filling was $1/r$, where $r\geqslant 1$ is a factor of aggregation [figure~\ref{fig3}~(a:I)].
  \item In the presence of a direct contact, the probability was $f$ for core-core contacts [figure~\ref{fig3}~(a:II)] and $1-f$ for core-shell contacts [figure~\ref{fig3}~(a:III)]. Here, $f$ is a solvation factor ($0<f\leqslant 1$) that accounts for core-core and core-shell interactions. The cases $f=1$ and $f=0$ correspond to the strong core-core and core-shell contacts, respectively.
\end{itemize}

The volume fraction of occupied sites $\varphi$ was determined as a ratio of the  number of filled sites $N$ and the total
number of sites $L_{x}\times L_{y}$. At an initial state, $L_{x}=L_{y}$ and $\varphi=\varphi_{\text i}$. In all calculations, the long lattice side was $L_{
x}=512$ and the initial concentration of conducting particles was  $\varphi_{\text i}=0.01$.

Figure~\ref{fig3}~(b) shows an example of evolution of aggregation patterns during the MC simulation of filtration-compression process. The data are
presented for a large factor of aggregation, $r=1000$, and weak solvation effects, $f=1$ (only core-core contacts are allowed).
Before compression, the size of the lattice was $L_{y}=L_{x}=512$. In the course of compression the lattice anisotropy $a=L_{x}/L_{y}$ $(a=\varphi/\varphi_{\text i})$ and the volume fraction of particles increase. At high values of $\varphi=0.501$, the electrical closure of contacts in vertical direction~$y$ was visually observed.

The MC model used makes it possible to account for the tendency of particles aggregation
(aggregation factor $r$) and formation of shells in the vicinity of conductivity particles (solvation factor, $f$). The value
of $r$ controls the degree of aggregation. At $r=1$, the aggregation is absent. However, it can be pronounced at $r\gg1$. The shells
have intermediate electrical conductivity. They are ``active'' and can also control aggregation. For example, for the case
$f\ll1$ (strong solvation), the deposition of a newcomer in the vicinity of the core of the previously deposited particle
is unlikely to take place and the core-shell contacts are mainly realized. In another limiting case  $f=1$ (weak solvation), the core-core contacts
are mainly realized.
At small values of $f$~($\ll 1$), the formation of checkerboard patterns with loose density is observed. In the limit of $f\to1$, the model predicts the formation of more compact and less spatially extended aggregates.

To simulate the filtration-compression process, the system was compressed along the vertical direction~$y$ by  sequentially removing the upper rows and redistributing the filled sites in these rows within the rest of the system. The fraction of occupied sites continuously increased with an increase of compression $\varphi=a\varphi_{\text i}$, where $a=L_{x}/L_{y}$ is a lattice anisotropy. The compression was stopped at $\varphi=1$ that corresponded to the final lattice anisotropy of $a=1/\varphi_{\text i}$. Periodical boundary conditions were used in the horizontal $x$ direction.

The Hoshen-Kopelman algorithm was used for labeling different clusters~\cite{Hoshen1976}. The value of the percolation threshold $\varphi_{\text c}$
corresponds to the minimum fraction of occupied sites at which an infinite cluster formed in the infinite lattice. Electrical
conductivity of the system was calculated using Frank and Lobb's algorithm~\cite{Frank1988}. This algorithm
utilizes a repeated application of a sequence of series, parallel and star-triangle $\text{Y}-\Delta$ transformations to the
square lattice bonds. The final result of this sequence of transformations is a reduction of a finite portion of the lattice to a single
bond that has the same conductance as the entire lattice portion. We used a scheme of four equivalent resistors (see,
e.g.~\cite{Cherkasova2010}) with high, $\sigma_{\text c}=10^{6}$, and low, $\sigma_{\text i}=1$, conductivity for the occupied and empty
sites, respectively.

\subsection{Statistical analysis}
The experiments were replicated $3{-}5$ times. In the MC calculations, the number of independent runs was  100. The mean values
and the standard deviations were calculated. The error bars in all the figures correspond to the confidence level 95\%.
The least square fitting of the experimental dependencies, determination of the fitting parameters and the coefficient of
determination, were provided using Table Curve 2D software (Jandel Scientific, USA).

\section{Results and discussion \label{sec:results}}

\subsection{Electrical conductivity of CNT-decane colloidal suspensions}
\begin{figure}[!b] 
\centering
\includegraphics [width=0.53\linewidth] {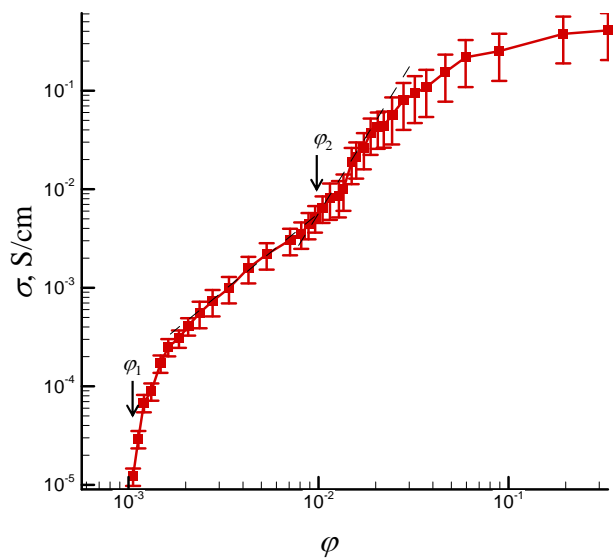}
\caption{(Color online) Electrical conductivity $\sigma$ versus volume concentration $\varphi$ of CNTs in decane. The value of $\sigma$
was measured using the vertical planar filtration-compression conductometric cell. The systems demonstrated the presence of two percolation thresholds at $\varphi_{1}\lesssim 10^{-3}$ and  $\varphi_{2}\approx 10^{-2}$.}
\label{fig4}
\end{figure}

Figure~\ref{fig4} presents experimental data on electrical conductivity of CNT-decane suspension, $\sigma$, versus the volume concentration of CNTs,
$\varphi$. The DP with thresholds at $\varphi_{1}\lesssim 10^{-3}$ and
$\varphi_{2}\approx 10^{-2}$ (estimated from inflection point in figure~\ref{fig4}) was observed. 
For large concentrations above $\varphi\approx 0.1$, the  $\sigma(\varphi)$ curve saturated.
Such a DP behavior of electrical conductivity can reflect the complexity of electrical contacts in the studied suspensions.
A similar behavior was previously observed in the liquid crystalline medium filled with CNTs ~\cite{Tomylko2015}. The simplest hypothesis explaining the DP behavior can be based on the model of shell-core structure
of CNT particles or their aggregates. In this model, the CNT particles with electrical conductivity $\sigma_{\text c}$ are surrounded by solvation shells
with electrical conductivity $\sigma_{\text s}$ that is intermediate between the conductivity of CNTs, $\sigma_{\text c}$, and continuous
medium, $\sigma_{\text i}$. The percolation threshold at a small concentration (at $\varphi_{1}\approx  10^{-3}$)
can be explained by interpenetration of the shells of loose CNT aggregates. With a further increase of $\varphi$ above
$\varphi_{1}$, the external pressure can cause transformation of large loose aggregates into the more compact aggregates.
Thus, the threshold at $\varphi_{2}\approx 10^{-2}$ can reflect the conducting path across the more compact aggregates with higher effective electrical conductivity. It is interesting to note that the saturation value of electrical conductivity $\sigma\approx 0.4$~S/cm for CNT-decane suspensions was still much smaller (approximately by one order of magnitude) compared to the experimentally measured value of $\sigma\approx 3$~S/cm for the pressed CNT-air system. It can reflect the formation of a tightly bound low conducting thin solvation layer of decane
near the surface of CNTs.

\subsection{Monte Carlo simulations}
\begin{figure}[!b] 
 \centering
\includegraphics[width=0.49\linewidth]{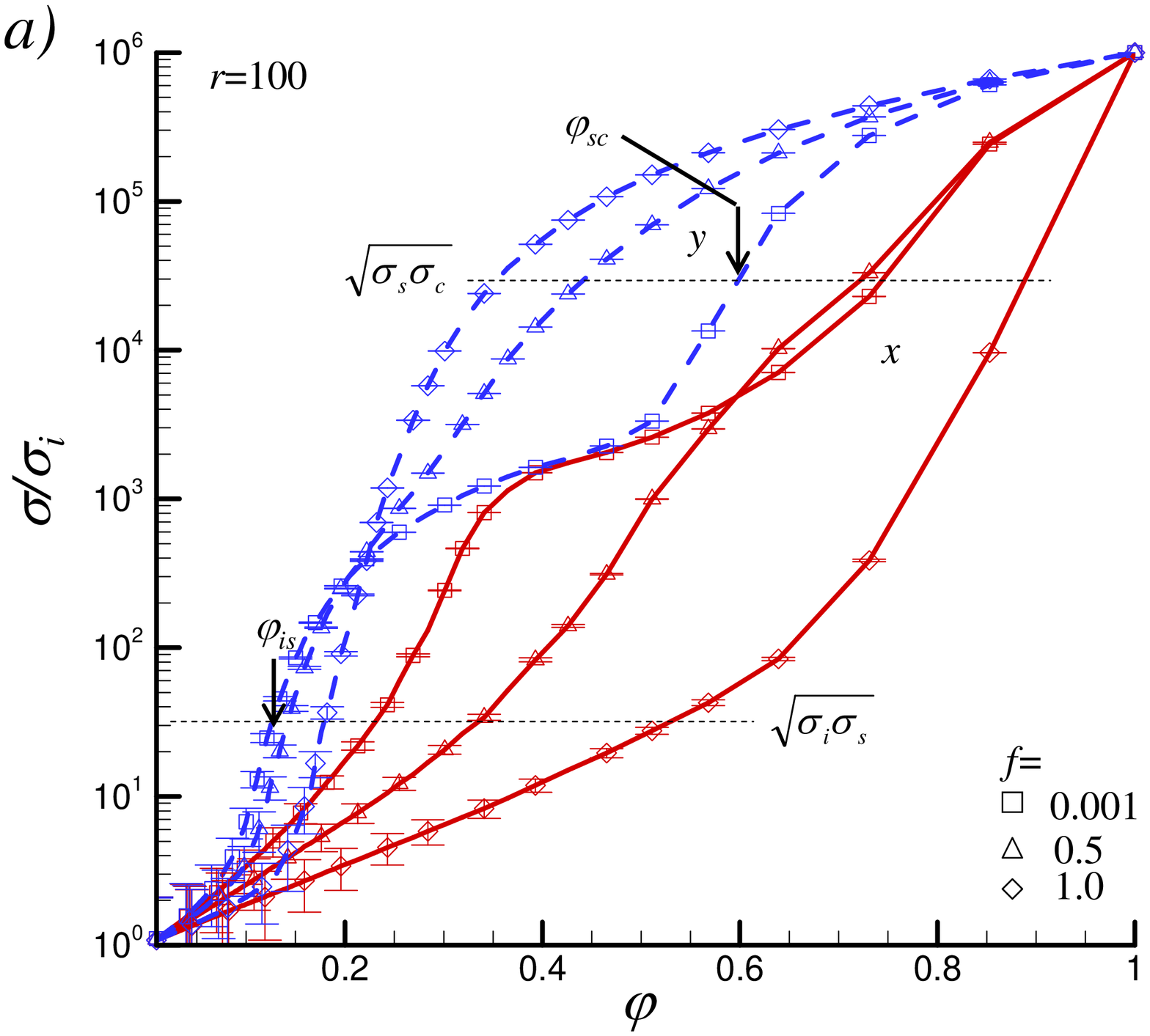}
\includegraphics[width=0.49\linewidth]{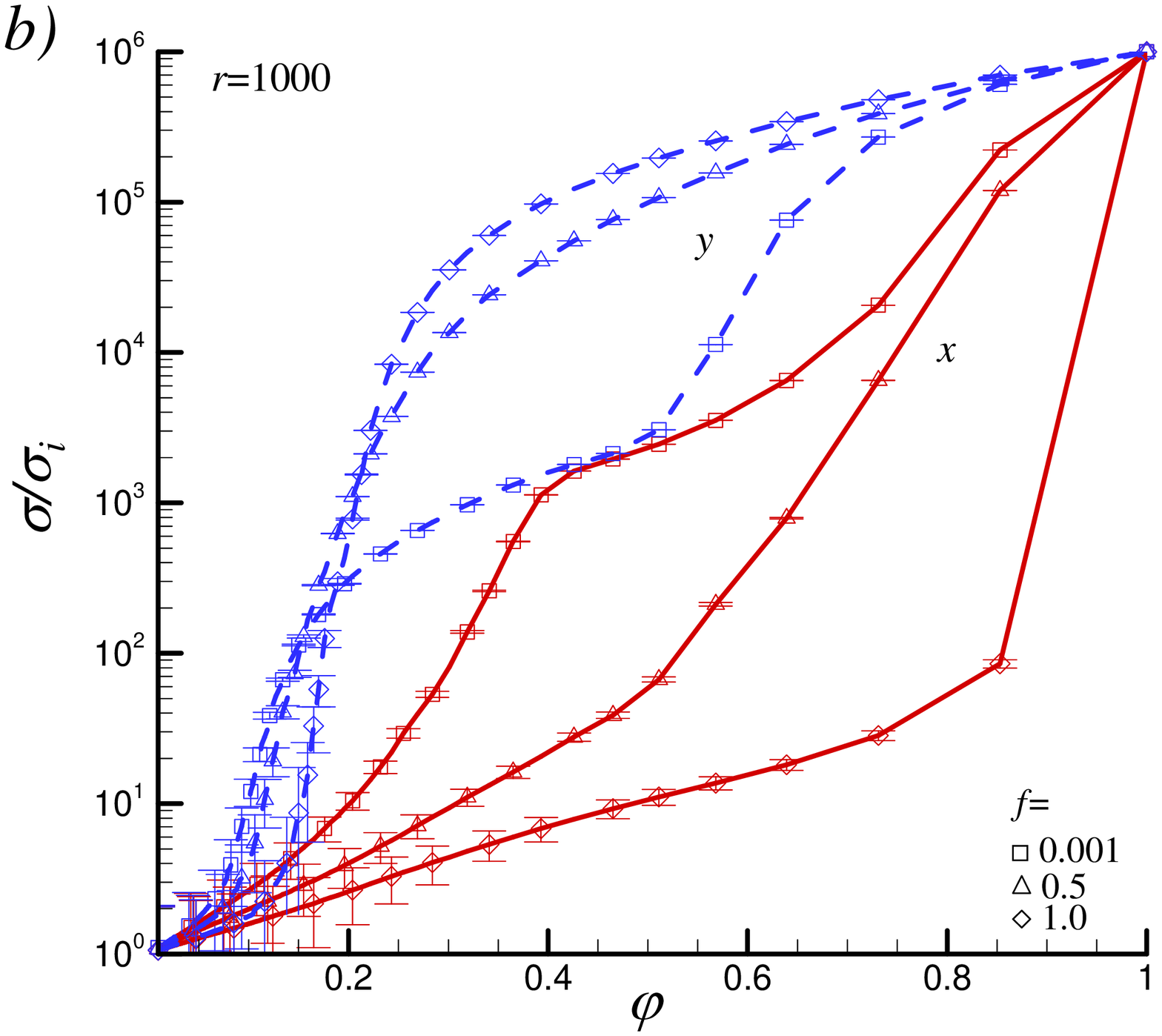}
\caption{(Color online) Dependencies of relative electrical conductivity $\sigma/\sigma_{\text i}$ versus concentration of particles $\varphi$
in horizontal ($x$, solid lines) and vertical ($y$, dashed lines) directions for different values of solvation factor, $f$,
and for the factor of aggregation $r=100$~(a) and $r=1000$~(b). \label{fig5}}
\end{figure}

Figure~\ref{fig5} presents examples of calculated dependencies of relative electrical conductivity $\sigma/\sigma_{\text i}$ versus concentration of particles $\varphi$ in horizontal ($x$, solid lines) and vertical ($y$, dashed lines) directions. The data are presented for $r=100$~(a) and $r=1000$~(b) and for different values of solvation factor, $f$.

The electrical conductivity along short vertical direction $y$, $\sigma_{y}$, always exceeded the value $\sigma_{x}$ along horizontal $x$ direction.
Anisotropy of electrical conductivity was absent at $r=1$ and became noticeable at large values of $r$~$(\gg1)$. The rapid grow of electrical conductivity can be explained by the formation of electrical closures in a vertical direction cased by the presence of large aggregates [figure~\ref{fig3}~(b)].
Moreover, at small values of $f$ ($f\ll1$, strong solvation), a clear DP behavior
with a sharp increase of electrical conductivity for the isolator-shell (at $\varphi=\varphi_{\text{is}}$) and shell-conductor (at $\varphi=\varphi_{\text{sc}}$) transitions was observed (figure~\ref{fig5}). However, at high values of $f$ ($f\rightarrow 1$, weak solvation), the electrical conductivity transitions were smeared over a wide interval of concentrations.
\begin{figure}[!t] 
\centering
\begin{overpic}[width=0.49\textwidth]{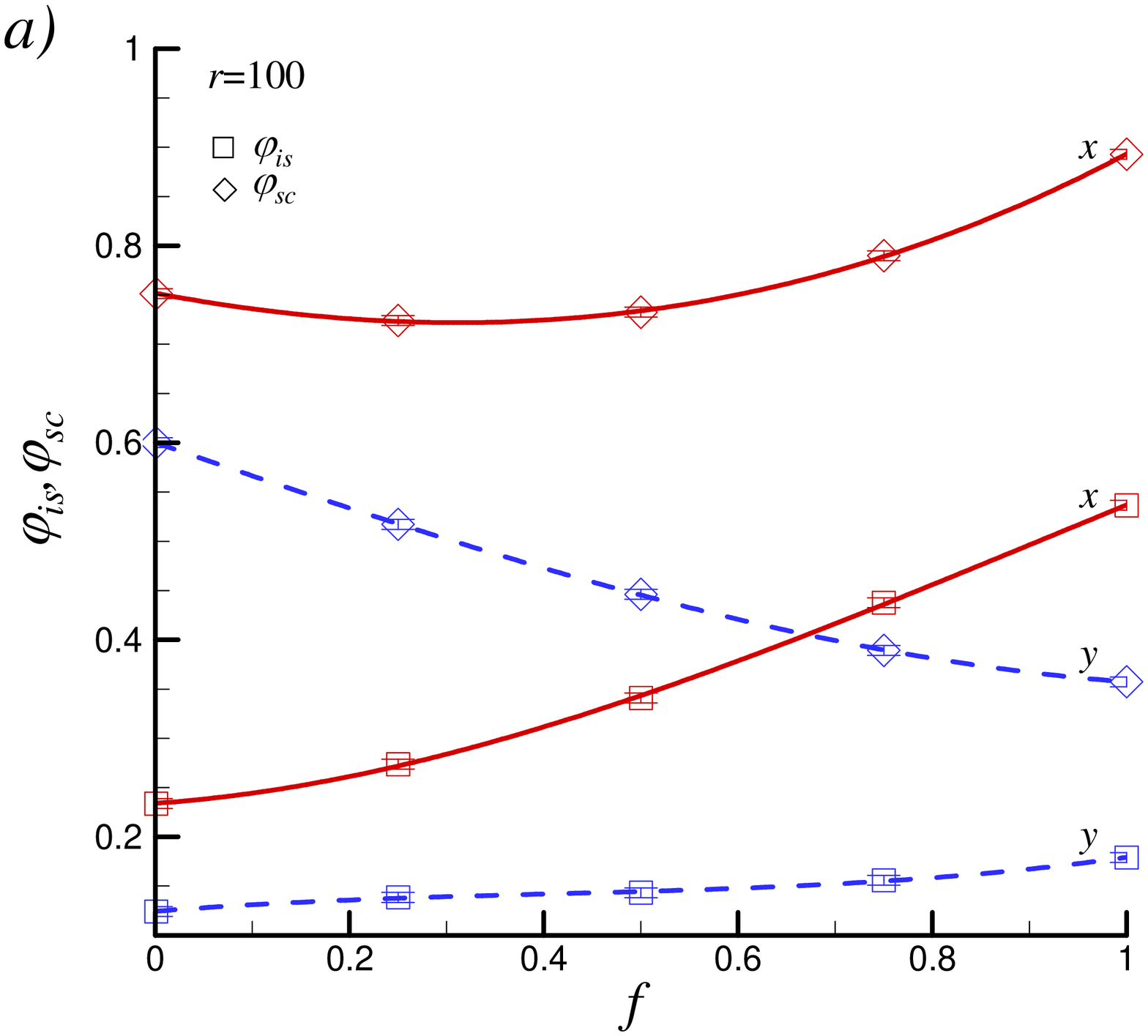}
\end{overpic}
\begin{overpic}[width=0.49\textwidth]{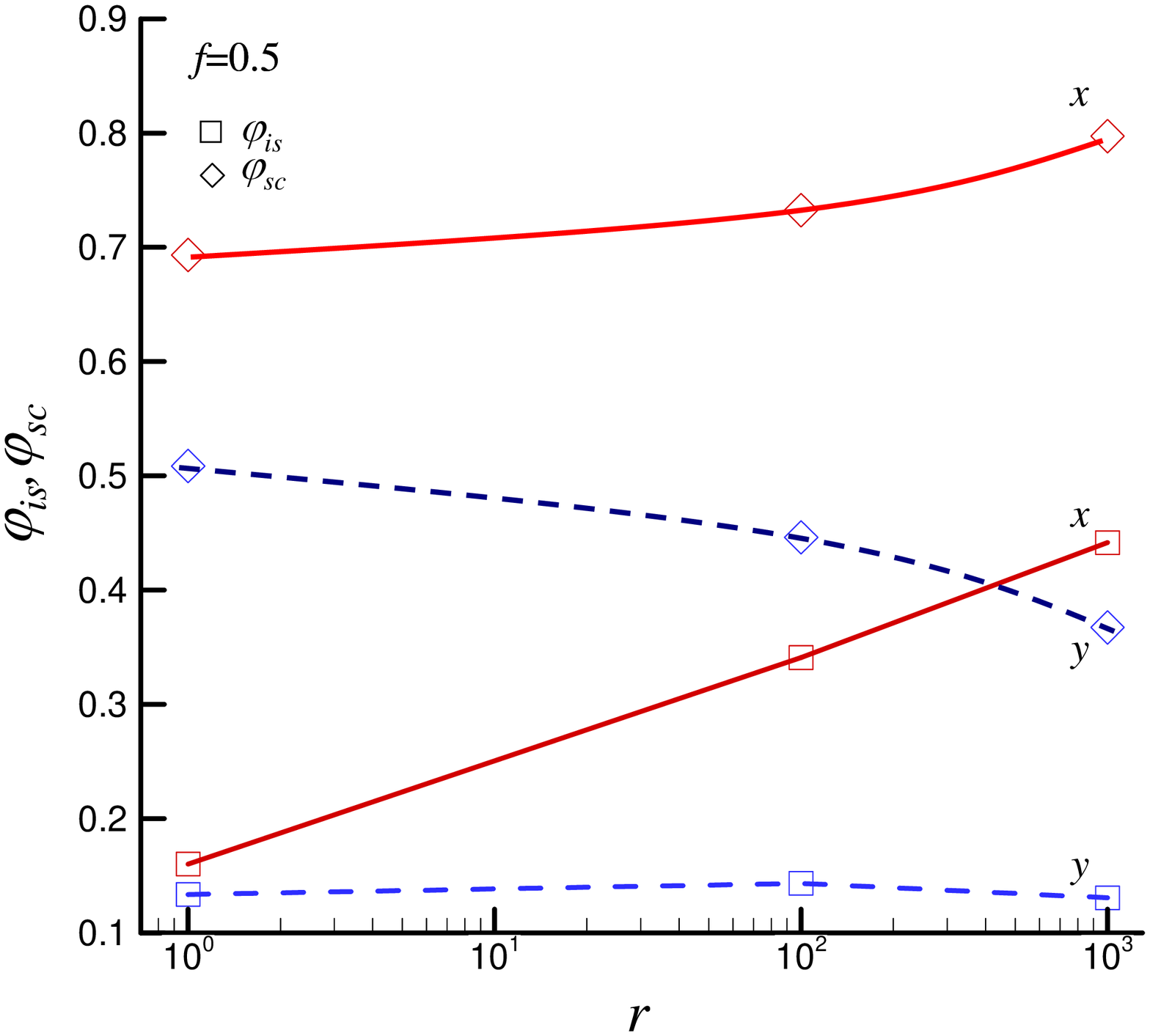}
\put(-1,85) {\large $b)$}
\end{overpic}
\caption{(Color online) Geometrical concentrations for the isolator-shell, $\varphi_{\text{is}}$, and shell-conductor, $\varphi_{\text{sc}}$, transitions
in horizontal ($x$, solid lines) and vertical ($y$, dashed lines) directions versus the factor of solvation $f$ at $r=100$~(a) and versus the factor of aggregation $r$ at $f=0.5$~(b). \label{fig6}}
\end{figure}
To characterize the percolation behavior in this work, the values $\varphi_{\text{is}}$ and $\varphi_{\text{sc}}$ along horizontal $x$ and vertical
$y$ directions were roughly estimated as geometrical concentrations that correspond to the mean geometrical conductivities $\sqrt{\sigma_{\text i}\sigma_{\text s}}$ and $\sqrt{\sigma_{\text s}\sigma_{\text c}}\,$, respectively [figure~\ref{fig5}~(a)]. Note that for the 2D systems with equal concentrations of the phases and their random distribution, the theory predicts a geometric conductivity at the percolation threshold \cite{Dykhne1971}. MC~simulation predicted
the geometrical concentration to be rather close to the percolation threshold~\cite{Tarasevich2016}.

Figure~\ref{fig6} presents these geometrical concentrations for the isolator-shell, $\varphi_{\text{is}}$, and shell-conductor, $\varphi_{\text{sc}}$, transitions in horizontal ($x$, solid lines) and vertical ($y$, dashed lines) directions versus the probability of the
face-to-face contact $f$ at a fixed factor of aggregation, $r=100$~(a) and versus a factor of aggregation $r$ at a fixed factor of solvation, $f=0.5$~(b).
In all cases we observed $\Delta\varphi_{\text{is}}=\varphi_{\text{is}}(x)-\varphi_{\text{is}}(y)>0$ and $\Delta\varphi_{\text{sc}}=\varphi_{\text{sc}}(x)-\varphi_{\text{sc}}(y)>0$, i.e., percolation was more enhanced in a vertical direction $y$ as compared with horizontal direction $x$.

At a fixed value of $r$, $\varphi_{\text{is}}$ (isolator-shell transition) increases for both $x$ and $y$ directions with an increase of $f$ [figure~\ref{fig6}~(a)].
However, the value $\varphi_{\text{sc}}$ (shell-conductor transition) increases for $x$ and decreases for  $y$ directions with an increase of $f$.
At a fixed value of $f$, the differences $\Delta\varphi_{\text{is}}$ and $\Delta\varphi_{\text{sc}}$ increase with an increase of the aggregation factor $r$ [figure~\ref{fig6}~(b)]. In horizontal direction $x$, the both geometrical concentrations $\varphi_{\text{is}}$ and $\varphi_{\text{sc}}$ increase with an increase of $r$. It reflects  suppression effect of aggregation on the percolation in $x$ direction. In vertical direction $y$, the opposite effect of aggregation is observed for shell-conductor transition, and the value of $\varphi_{\text{sc}}$ decreases with an increase of $r$. However, the aggregation has practically no effect on geometrical concentrations for the isolator-shell, $\varphi_{\text{is}}$, transition.

\section{Conclusions and final remarks \label{sec:conclusion}}

The experimental data obtained for electrical conductivity, $\sigma$, of colloidal CNT-decane suspension measured in the
planar filtration-compression conductometric cell evidenced the presence of two percolation thresholds at
$\varphi_{1}\lesssim 10^{-3}$  and  $\varphi_{2}\approx 10^{-2}$. The experimentally observed DP can be explained on the basis of shell-core model of CNT particles or their aggregates. The percolation threshold at $\varphi_{1}$ and  $\varphi_{2}$
can reflect the interpenetration of low conducting shells of loose CNT aggregates and percolation across the more compact
conducting aggregates, respectively.

The MC model proposed accounted for the core-shell structure of conducting particles, the tendency of particle aggregation,
the formation of solvation shells, and an elongated geometry of the conductometric cell. The studies revealed DP
transitions that correspond to the percolation through the shells and cores. The MC data also demonstrated a noticeable
impact of particle aggregation on anisotropy in $\sigma(\varphi)$ dependencies measured along different directions in the
conductometric cell. Simulation data predicted a rather complex behavior for electrical conductivity curves in dependence on aggregation, $r$, and
solvation, $f$, factors. The impact of $r$ and $f$ on the percolation behavior can be explained accounting for the differences in the structure of
aggregates. An easier percolation along the shorter vertical axis $y$ for strongly aggregated systems is expected.
The impact of solvation factor $f$ on the percolation behavior can reflect the changes in the compactness of aggregates. For strong solvation effects, $f\ll1$, loose aggregates with small density were formed while  for weak solvation effects, $f\approx1$, the aggregates became more compact and less spatially extended. However, quantitative comparison between experiments and the present model is difficult due to fairly artificial assumptions of the model accounting for the tendency of particles aggregation and formation of low conducting shells in the vicinity of conductivity particles. Similar difficulties have been recently stated in the studies of percolation of disordered segregated composites~\cite{Johner2009}. Future work should consider a more realistic model with account for the tunneling transport between conducting particles and an elongated shape of CNTs.

\section*{Acknowledgements}
This work was partially funded by the National Academy of Sciences of Ukraine, Projects No. 2.16.1.4 and No. 43/17-H.

\ukrainianpart
\title{Двосхідцева перколяція в агрегованих системах}
\author{М.І. Лебовка\refaddr{label1}, 
Л.~Булавін\refaddr{label2},
В.~Ковальчук\refaddr{label2}, 
І.~Мельник\refaddr{label2}, 
K.~Репнін\refaddr{label2} 
}
\addresses{
\addr{label1} Інститут біоколоїдної хімії ім. Ф.Д. Овчаренка НАН України, \\ бульв. акад. Вернадського, 42, 03142 Київ, Україна
\addr{label2} Фізичний факультет, Київський національний університет ім. Тараса Шевченка, \\ пр. акад. Глушкова, 2, 03127 Київ, Україна
}

\makeukrtitle
\begin{abstract}
Двосхідцева перколяційна поведінка в агрегованих системах була досліджена експериментально за допомогою моделювання методом Монте Карло. В експериментальних дослідженнях були проведені вимірювання електричної провідності, $\sigma$, колоїдних суспензій багатошарових вуглецевих нанотрубок у декані. Вимірювання проводились в планарній фільтраційній компресійній кондуктометричній комірці при механічному віджиманні рідини з суспензії. При віджиманні відстань між електродами зменшувалась, і об'ємна концентрація нанотрубок в $\varphi$ в декані збільшувалася (від $10^{-3}$ до $\approx 0.3$\%~v/v). Спостерігалися два перколяційні переходи при $\varphi_{1}\lesssim 10^{-3}$ і $\varphi_{2}\approx 10^{-2}$, які могли відповідно відображати взаємопроникнення рихлих агрегатів нанотрубок і перколяцію по компактних провідних агрегатах. Обчислювальна модель Монте Карло враховувала наявність у провідних частинок або їх агрегатів структури типу ядро-оболонка, тенденцію частинок до агрегації, утворення сольватних шарів на поверхні частинок і наявність подовженої геометрії кондуктометричної комірки. Комп'ютерна модель дозволила виявити наявність двох розмазаних перколяційних переходів в концентраційних залежностях $\sigma(\varphi)$, які відповідали перколяції по оболонках і ядрах. Спостерігався значний вплив агрегації частинок на анізотропію електропровідності в різних напрямках кондуктометричної комірки.
\keywords{багатошарові вуглецеві нанотрубки, колоїдні суспензії, анізотропія електричної провідності, двосхідцевий перколяційний перехід}
\end{abstract}
\end{document}